\documentclass[pra,twocolumn,showpacs,superscriptaddress,preprintnumbers,amsmath,amssymb,aps]{revtex4-1}
\usepackage{mathrsfs}
\usepackage{amsmath}
\usepackage{amssymb}
\usepackage{graphicx}
\usepackage{dcolumn}
\usepackage{bm}
\usepackage{subfigure}
\usepackage{color}
\usepackage{ifpdf}
\usepackage{epstopdf}
\usepackage{CJK}
\usepackage{verbatim}
\usepackage{bbm}
\usepackage{multirow}
\usepackage{booktabs}
\usepackage{color}
\usepackage[
pdfstartview=FitH,%
bookmarks=true,%
bookmarksopen=true,%
breaklinks=true,%
colorlinks=true,%
linkcolor=blue,anchorcolor=blue,%
citecolor=blue,filecolor=blue,%
menucolor=blue,pagecolor=blue,%
urlcolor=blue]{hyperref}
\usepackage{float}
\usepackage[toc,page,title,titletoc,header]{appendix}
\usepackage{ragged2e}
\usepackage{lipsum}

\usepackage{array}
\usepackage{booktabs}

\usepackage{color}

\def\be{\begin{equation}}
	\def\ee{\end{equation}}
\def\bea{\begin{eqnarray}}
	\def\eea{\end{eqnarray}}

\begin{document}

\title{Temporal Talbot interferometer of strongly interacting molecular Bose-Einstein condensate}
\author{Fansu Wei}
\affiliation{State Key Laboratory of Advanced Optical Communication Systems and Networks, School of Electronics, Peking University, Beijing 100871, China}
\author{Zhengxi Zhang}
\affiliation{State Key Laboratory of Advanced Optical Communication Systems and Networks, School of Electronics, Peking University, Beijing 100871, China}
\author{Yuying Chen}
\affiliation {School of Physics and Electronics Engineering, Shanxi University, Taiyuan 030006, China}
\author{Hongmian Shui}
\affiliation{State Key Laboratory of Advanced Optical Communication Systems and Networks, School of Electronics, Peking University, Beijing 100871, China}
\affiliation{Institute of Carbon-based Thin Film Electronics, Peking University, Shanxi, Taiyuan 030012, China}
\author{Yun Liang}
\affiliation{State Key Laboratory of Advanced Optical Communication Systems and Networks, School of Electronics, Peking University, Beijing 100871, China}
\author{Chen Li}\email{chen.li@tuwien.ac.at}
\affiliation{Vienna Center for Quantum Science and Technology, Atominstitut, TU Wien, Stadionallee 2, 1020 Vienna, Austria}
\author{Xiaoji Zhou}\email{xjzhou@pku.edu.cn}
\affiliation{State Key Laboratory of Advanced Optical Communication Systems and Networks, School of Electronics, Peking University, Beijing 100871, China}
\affiliation{Institute of Carbon-based Thin Film Electronics, Peking University, Shanxi, Taiyuan 030012, China}
\affiliation{Institute of Advanced Functional Materials and Devices, Shanxi University, Taiyuan 030031, China}
\date{\today}

	\begin{abstract}
		
		Talbot interferometer, as a periodic reproduction of momentum distribution in the time domain, finds significant applications in multiple research. The inter-particle interactions during the diffraction and interference process introduce numerous many-body physics problems, leading to unconventional interference characteristics. This work investigates both experimentally and theoretically the influence of interaction in a Talbot interferometer with a $^{6}\rm Li_2$ molecular Bose-Einstein condensate in a one-dimensional optical lattice, with interaction strength directly tunable via magnetic Feshbach resonance. A clear dependence of the period and amplitude of signal revivals on the interaction strength can be observed. While interactions increase the decay rate of the signal and advance the revivals, we find that over a wide range of interactions, the Talbot interferometer remains highly effective over a certain evolutionary timescale, including the case of fractional Talbot interference. This work provides insight into the interplay between interaction and the coherence properties of a temporal Talbot interference in optical lattices, paving the way for research into quantum interference in strongly interacting systems.
	\end{abstract}
 \maketitle

\begin{figure*}[htp]
    \includegraphics[width=0.8\textwidth]{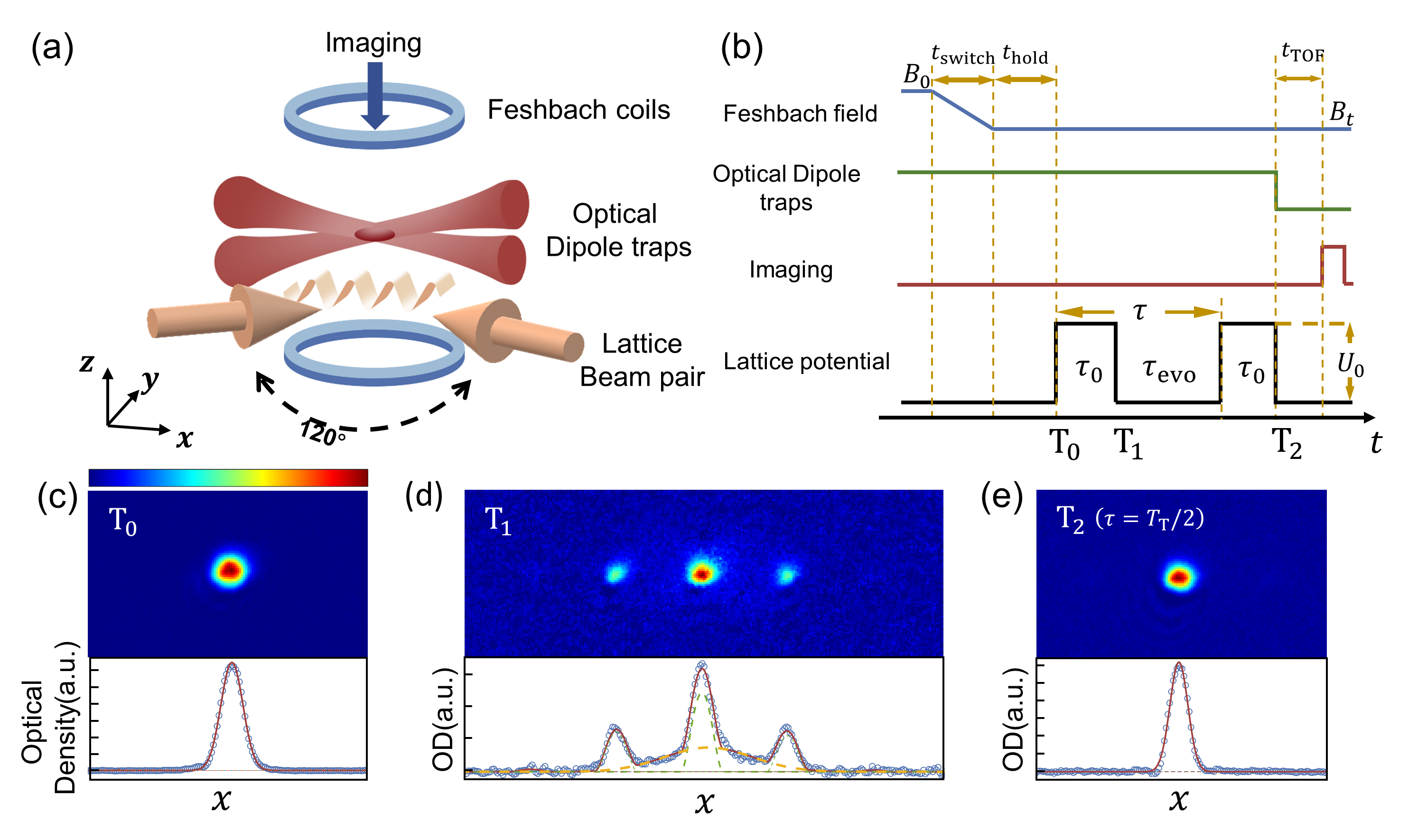}
    \caption{
		{\bf Schematic of the experimental system}.
		(a) Schematic of the experimental system. Feshbach mBECs are trapped in a pair of crossover dipole traps. The blue circles in the $x-y$ plane represent the Feshbach magnetic coils and the red Gaussian beams along the $x$-axis mark the crossed optical dipole trap. The two orange arrows in the $x-y$ plane stand for the lattice beams with a crossing angle of $120^{\circ}$. The blue arrow shows the imaging direction, which is perpendicular to the $x-y$ plane. (b) A typical experimental time sequence. The black line represents the optical lattice, and the two pulses with duration of $\tau_0$ before and after $\tau_{evo}$ are Talbot pulses. The interference time $\tau = \tau_{\rm evo} + \tau_{0}$. The blue line and green line represent the strength of the Feshbach field and optical dipole trap, respectively. The Feshbach field is switched from $B_0$ to $B_t$ in $t_{\rm switch}$ and kept for a period of $t_{\rm hold}$. The optical dipole trap remains constant and is turned off together with the second lattice pulse. The red line represents the absorption imaging pulse, which is applied after the time-of-flight process with time $t_{\rm TOF}$. (c) $\rm T_0$: mBEC. (d) $\rm T_1$: the scattering pattern with pulse $U_0 = 50E_r$ and $t = 0.7\mu \rm s$. (e) $\rm T_2$: the interference pattern with $\tau_{\rm evo} = T_{\rm T}/2$. (top) Raw image after $3$ ms TOF ($10$ images averaged) for $a_{dd} = 865 a_0$ and (bottom) the corresponding bi-mode fitting result for distinguishing between condensed and non-condensed particles.}
	\label{fig:experimentalSetup}
\end{figure*}
	\section{Introduction}
    The Talbot effect constitutes a near-field interference phenomenon wherein a periodic pattern undergoes self-imaging after passing through a diffraction grating \cite{Talbot1836}. The applicability of the paraxial approximation is pivotal to this near-field interference effect, with the periodic revivals stemming from phase coherence across adjacent grating slits \cite{LOHMANN1971413,Keren:85}. The Talbot effect and its variants have been harnessed in diverse domains, including X-ray imaging \cite{Momose_2003, Momose_2006}, waveguide arrays \cite{1548914, Chen2015DiscreteTE}, and plasmonics \cite{Dennis2007ThePT,Li2011TheTE}, influencing classical \cite{Rayleigh1881}, nonlinear \cite{PhysRevLett.104.183901}, and quantum optics research \cite{Wen:13} significantly.
    
    Observations of the Talbot effect extend to atomic and molecular matter-wave interference \cite{PhysRevLett.83.5407, Mark_2011, PhysRevA.88.013603}. Ultra-cold gases, with their exceptional controllability and advanced measuring approaches \cite{Schafer2020}, offer a robust experimental framework for exploring novel quantum states \cite{PhysRevLett.126.035301}, orbit-based quantum simulations \cite{Bloch2012,PhysRevLett.121.265301,PhysRevA.107.023303}, quantum computing \cite{PhysRevA.84.032322,PhysRevA.104.L060601}, precision metrology\cite{GUO20222291,Dong_2021}, and macroscopic matter-wave interference \cite{Grond_2010,Bloch_2005,Hu2018}. Lattice pulses can impart varying momentum distributions to particles. Utilizing a pair of such pulses as gratings creates a temporal Talbot interferometer, enabling observation of periodic revivals with adjustable intervals \cite{Xiong_2013}, measurable via time-of-flight imaging \cite{PhysRevLett.101.155303}.

    In interference experiments, efforts typically focus on minimizing interaction-induced coherence perturbations \cite{Hofferberth2008}. Such interactions often result in pronounced collisions, evident in the s-wave scattering halos of atomic clouds expanding from an optical lattice \cite{PhysRevD.78.042003, PhysRevA.92.013616}. Nevertheless, interactions are an inherent variable and necessitate quantitative analysis for appropriate adjustment.

    Heretofore, the impact of interactions on Talbot interferometry has been cursorily addressed only in Ref. \cite{Santra2017}, with a preliminary assessment of the influence of tunable on-site interactions presented in Ref. \cite{PhysRevA.100.063613}. Yet experimental results elucidating Talbot interference within the context of direct interaction modulation remain unreported.
    Here, we study both experimentally and theoretically the effects of interactions within a Talbot interferometer, utilizing a $^{6}\rm Li_2$ molecular Bose-Einstein condensate(mBEC) in a one-dimensional optical lattice. The interaction strength is precisely controlled via magnetic Feshbach resonance \cite{RevModPhys.82.1225}. We observe the dynamics of Talbot signal revivals across diverse interaction conditions and find a distinct relationship between interactions and the periodicity and intensity of the signal revivals. In addition, interference behaviors of the fractional Talbot effect \cite{Ullah2012} are also observed in the presence of strong interactions. Numerical simulations were conducted to compare with experimental data and aid in elucidating the underlying physical mechanisms.

    
    This paper is organized as follows. In Sec. \ref{sec:experimentImplementation}, we describe our experimental procedure and the implementation method of the Talbot interferometer with different interactions. In Sec. \ref{sec:theoretic}, the theoretical model for the Talbot interferometer in a 1D optical lattice with strong interaction strength is described. In Sec. \ref{sec:decayResult} and Sec. \ref{sec:shiftResult}, we present the experimental results of Talbot signals decay and Talbot revivals shift under different interaction strengths, respectively. The experimental results of the fractional Talbot effect in the presence of interactions are described in Sec. \ref{sec:fractional}. Finally, we give the conclusion in Sec. \ref{sec:Conclusion}.

    \section{Temporal Talbot interferometer}\label{sec:experimentImplementation}
    Our experiments are performed with BECs of $^{6}\rm Li$ Feshbach molecules (refer to Appendix \ref{sec:A}). In each cycle we prepare a two-state mixture of Lithium atoms in the lowest hyperfine states $|F = 1/2,m_F  = 1/2\rangle $($|1\rangle $) and $|F = 1/2,m_F  = -1/2\rangle $($|2\rangle $). Then, through evaporation cooling in a cross-beam dipole trap at the unitary limit ($832$ Gauss) of the Feshbach resonance\cite{PhysRevLett.94.103201} and switched to the BEC side ($670-750$ Gauss in our experiment), we obtain the mBECs of $\sim 20000$ $^{6}\rm Li_2$ Feshbach molecules\cite{doi:10.1126/science.1093280}. By controlling the magnetic field we can tune the $s$-wave scattering length between molecules, which is given by $a_{dd} = 0.6a_{12}$\cite{PhysRevLett.93.090404}, where $a_{12}$ is the $s$-wave scattering length between atoms in states $|1\rangle $ and $|2\rangle $.

    Fig.\ref{fig:experimentalSetup}(a) shows the schematic of the experimental set-up. The mBECs are confined in a trapping potential formed by a pair of far-red-detuned lasers in a vertical plane with a $30^{\circ}$ to each other. A pair of hollow electric coils produce the Feshbach Resonance magnetic field. The combined trapping frequencies are  $(\omega_x,\omega_y,\omega_z ) = 2\pi \times (20,128,132)$ Hz, where $x$-axis refers to the horizontal direction of the plane trapping beams located, $y$ the other horizontal direction, and $z$ the vertical direction.  
    We use two beams of $\lambda = 1064\ \rm nm$ lasers, mutually separated by $\theta = 120^{\circ}$, to form a one-dimensional lattice potential $U(x) = U_0  \mathrm{cos}^2(\pi x/D)$ in the horizontal direction, where $U_0$ is the lattice potential depth, $D = \lambda/2\rm sin(\mathit{\theta}/2) = 614.3\ \rm nm$ the lattice spatial period. The lattice laser beams are focused to beam waists of $(w_{\rm Horizontal},w_{\rm Vertical}) = (250,110)\ \mathrm{\rm\mu m}$. The beams are large compared to the size of the mBEC, and hence the lattice potential depth is approximately uniform across the cloud. The two lattice laser beams are derived from the same laser source, intensity controlled by an acousto-optic modulator (AOM), and split by a polarizing beam splitter to about 50:50, on-off controlled by two other AOMs synchronously. The characteristic lattice energy is $E_r = \hbar^2 \mathit{k}^2/2\mathit{m} $, where $k = \pi/D$ and $m$ is the mass of a lithium molecule ($^{6}\rm Li_2$).

    The experimental time sequence is presented in Fig.\ref{fig:experimentalSetup}(b). After the completion of evaporation cooling, the Feshbach magnetic field is adiabatically ($500 \ \mathrm{Gauss/s}$) ramped from $B_0$ to $B_{\rm t}$ within $t_{\rm switch}$ and kept for an additional duration of $t_{\rm hold}=100$ ms to stabilize. The lattice potential is then pulsed on twice with a variable evolution interval $\tau_{\rm evo}$ between. The Talbot interference time $\tau$ is defined by the sum of the evolution time $\tau_{\rm evo}$ and the duration of the first pulse $\tau_{0}$. The optical dipole traps remain unchanged throughout the experiment, so a well-defined geometry and interaction energy are maintained during the scattering and interference process. After the second pulse, detection is performed by absorption imaging with $t_{\rm TOF}=3$ ms TOF (time-of-flight). Fig. \ref{fig:experimentalSetup}(c) depicts the absorption image after $3$ ms TOF before the first pulse ($\rm T_0$) as mBEC. And Fig. \ref{fig:experimentalSetup}(d) depict the $3$ ms TOF absorption imaging results taken after the first pulse ($\rm T_1$), showing the momentum distribution of particles after Kaptiza-Dirac(KD) scattering, which refers to two-photon scattering processes when particles move through a lattice\cite{GUPTA2001479}. The $3$ ms TOF image after the second pulse ($\rm T_2$) with an interference time of $\tau = T_{\rm T}/2$ is shown in Fig. \ref{fig:experimentalSetup}(e).

\begin{figure}
	\includegraphics[width=0.45\textwidth]{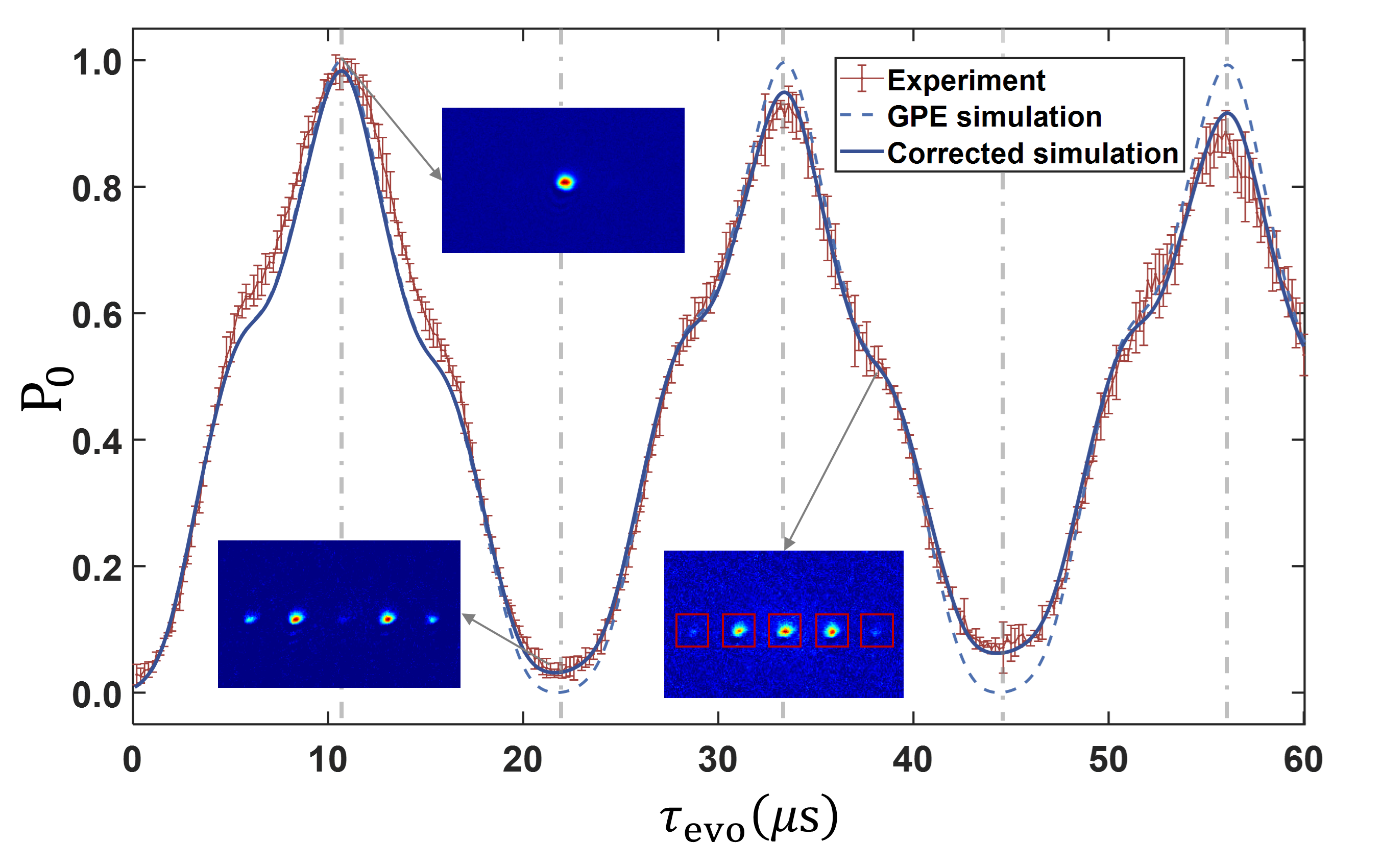}
	\caption{
		{\bf Temporal Talbot interferometer of strongly interacting mBEC}.
		The plot shows the proportion of the $0 \hbar k$ particles over the total number of particles in the absorption image for different free evolution times between two pulses of a 1D optical lattice for $a_{dd} = 865 a_{0}$. The red dots are experimental results, and the error bar shows the standard error of five measurements. The blue dashed and solid lines are the theoretical results without damping and with exponential decay correction, respectively. Raw plots of three different nodes during oscillation are presented in the insets. The red box indicates the molecular statistical range of each momentum mode. The positions of $T_{\rm T}/2$ multiples are indicated by gray dotted lines.
 }
	\label{fig:duration}
\end{figure}

    In the experiment, the lattice depth of the pulses, which is calibrated by KD scattering, is set to $U_0 = 50 E_r$ and the duration of both pulses is $\tau_0 = 0.7\mathrm{\rm\mu s}$. The pulse time is short enough to keep within Raman-Nath regime $t\sqrt{E_r U_0} /\hbar <1$ \cite{Raman1935} so that the particles remain approximately stationary during the lattice pulse and the interference process. In Fig. \ref{fig:duration}, the relative population $\rm P_0$ as a function of the varied interval $\tau_{\rm evo}$ for $a_{dd} = 865 a_{0}$ is shown by the red dots, with $\tau_{\rm evo}$ changed every 0.2 $\rm\mu s$ up to $60 \rm\mu s$. $\rm P_0$ $= N_0/N_p$, in which $N_0$ is the particle number of $0 \hbar k$ modes, $N_p$ the total number of particles. 
    
    Here, we introduce Method 1 for obtaining Talbot signals (also the traditional one). $N_0$ and $N_p$ are obtained by integrating the number of particles in square regions near each momentum peak, as indicated by the red box in the inset. The side length of boxes is $84 \mathrm{\mu m}$, equivalent to the range of $\pm 0.5\hbar k$ for each momentum mode.
    
    The Talbot time in our system is $T_{\rm T} \simeq 22.5 \mathrm{\rm\mu s}$, while the theoretical value is $T_{\rm theoretic}= h/4E_r = 22.7 \mathrm{\rm\mu s}$. The two essentially stay the same, with the slight deviation attributed to a minute discrepancy in the angle of the lattice beams. The positions of $T_{\rm T}/2$ multiples are indicated by gray dotted lines in Fig. \ref{fig:duration}. When there is no significant momentum broadening in the initial state or considerable interaction, $\rm P_0$ will become maximal close to $1$ at odd multiples of the half Talbot time $T_{\rm T}/2$, minimal nearly $0$ at even multiples of $T_{\rm T}/2$, just as the theoretical simulation result shows (blue dashed line). However, in the case of strong interactions, the decay of the phase correlations cannot be ignored and is reflected in an exponential decay. We apply an exponential correction to the simulation result and find it aligns well with the experimental data.

    \section{Theoretical analysis}\label{sec:theoretic}

    To further understand the experiment, we conduct a theoretical analysis of the temporal Talbot interferometer under various interactions. The simulations employ a mean-field approach based on the Gross-Pitaevskii equation (GPE),
    \begin{equation}
    \label{eqGPE}
    i \hbar \frac{\partial \Psi}{\partial t} = [-\frac{\hbar^2}{2m} \frac{\partial^2}{\partial x^2} +\frac{1}{2} m\omega_x^2 x^2 +U(x) +g |\Psi|^2] \Psi,
    \end{equation}
    where $m$ is the mass of a molecule $\rm ^{6}Li_2$, $\omega_x$ is the combined trapping frequency, $U(x)$ is the lattice potential and $g \sim a_{s}$ the effective interaction constant. In our system, the dynamics in the elongated direction are more relevant, while the transverse directions can be integrated out, leading to an effective 1D description with the 1D interaction parameter, $g_{1D}= 16 \hbar^2 a_{dd}/3 m R_{TFr}^2$. The $R_{TFr}$ is the Thomas-Fermi radius along the radial directions of the 3D mBEC and is calculated with the radial combined trapping frequencies ($\omega_y$, $\omega_z$) which are calibrated at a set of scattering lengths.

    We implement simulations of the Talbot interferometer with different interaction strength (scattering length $a_{dd}$), different optical lattice pulses (lattice depth $U_0$ and pulse duration $\tau_0$), and different characteristic lattice energy $E_r$. The simulation results under the same conditions as the experiment are presented in Fig. \ref{fig:duration} and \ref{fig:shift} for comparison, while the others are presented in Appendix. \ref{sec:B} to show the overall trend. It demonstrates that the effect of interaction on Talbot interference can be approximately determined by the values of $g/E_r$ and $U_0 \tau_0$. 
    
    When $g/ E_r$ is small, the signal curve changes quite little compared to the situation without interaction. For larger $g/E_r$, the whole curve compresses towards $t=0$, and the peaks exhibit a negative temporal shift as well as an amplitude decay. From an analytical perspective, the interaction term in \eqref{eqGPE} can be approximately treated as a shallow lattice term with a position-dependent amplitude. It induces an extra phase in the free evolution stage like an ordinary optical lattice, resulting in a positive correction on $E_r$. Therefore, $T_{\rm theoretic}= h/4E_r$ will decrease and cause a negative temporal shift on peaks. Meanwhile, a higher proportion of high-momentum modes is produced, strengthening the collision and decoherence among particles to make a signal damping. Both of these effects are caused by interactions during the evolution stage, rather than the lattice pulse stage.

    Besides, as $U_0\tau_0$ increases, a greater number of higher-order momentum modes emerge, accompanied by the appearance of sub-maximal peaks in the time domain, other than the principal maxima of the Talbot signal. The phenomenon is commonly referred to as the fractional Talbot effect, and there are some different behaviors under strong interactions.


    \section{Talbot signal damping under different interactions}\label{sec:decayResult}
 \begin{figure}
    \includegraphics[width=0.45\textwidth]{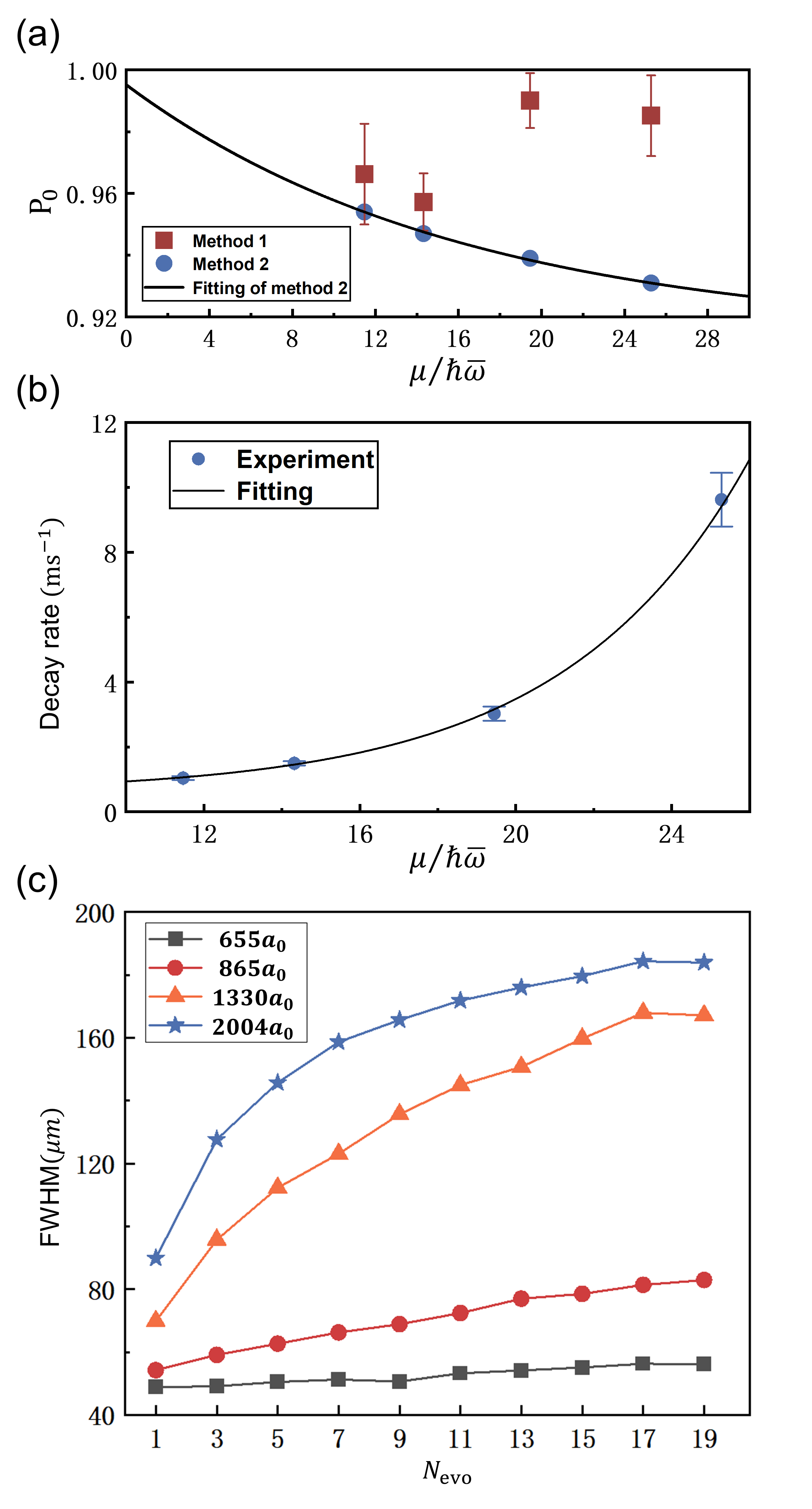}
    \caption{
    {\bf The damping of the Talbot signals under different interactions}.
    (a) The first peak of $\rm P_0$ under different interactions. The red square dots with error bars are the results of Method 1, and the blue circular dots are the results of Method 2, fitted with the black line. (b) The decay rate of the Talbot signals under different interactions. (c) The variation of the full width at half maximum (FWHM) of the molecular cluster with momentum $0\hbar k$, under different interactions, as a function of the interference period. The length corresponding to $2\hbar k$ in the momentum space is $168\ \mathrm{\mu m}$.}
    \label{fig:decay}
\end{figure}

    To experimentally delineate the effect of interaction on the damping of temporal Talbot signals, we evaluate the decay rates for varying interaction strengths modulated via Feshbach resonance. The scattering lengths in experiment are designated as $a_{dd} = 655 a_{0}$, $865 a_{0}$, $1330 a_{0}$, $2004 a_{0}$. In our experiment, there is a notable reduction in particle count as $a_{dd}$ decreases. To eliminate this disturbance, we utilize $\mu/\hbar \bar{\omega } \sim (N_p a_{dd})^{5/2}$ as the standard measure of interaction strength. 
    
    Initially, we employed Method 1 to acquire the signals. At $\tau=T_{\rm T}/2$, the first peak of $\rm P_0$ showed no significant difference under varying interactions (red dots in Fig. \ref{fig:decay}(a)). We find that the damping effect does not increase monotonically with interaction strength, it diminishes after reaching a peak rather (refer to Appendix \ref{sec:C} for details). Analysis of the raw data reveals that strong interactions generate numerous thermal particles near the $0 \hbar k$ momentum mode, originating from collisions both during the interference phase and when different momentum components segregate during the TOF stage. These particles cause distortions in the Talbot signals and likely do not contribute to the oscillatory interference process. This influence becomes more pronounced when $\mu/\hbar \bar{\omega }\ge 20$.

    Adopting the method from Ref. \cite{SciPostPhys2022}, we introduce Method 2 to refine the extraction of Talbot signals. A bi-modal fit is used to separate the condensed and non-condensed fractions, with the latter being excluded from the analysis of the Talbot signal (Fig. \ref{fig:experimentalSetup}(d) bottom panel).

    The signals from Method 2 at $\tau=T_{\rm T}/2$, represented by blue dots in Fig. \ref{fig:decay}(a), correspond to the averaged fitted results from 25 datasets. Contrary to the results from Method 1, a significant reduction in the amplitude of the first revival correlated with increased interaction strength is noted, implying that interactions induce damping during the pulse, initial evolution, and TOF stages. However, the effect is minor as $\rm P_0$ remains above $0.9$. Furthermore, the fitting of these points yields $\rm P_0(\tau=T_{\rm T}/2) \sim 1$ in the absence of interaction. Applying this method, we chart the decay across the first 10 revivals for four distinct interaction strengths to derive decay rates linearly (Appendix \ref{sec:C}). The absolute values of the slopes are considered as the decay rates, and represented by blue dots in Fig. \ref{fig:decay}(b). Upon fitting an exponential curve, we notice a significant increase in signal damping as the interaction strength increases. This indicates a significant impact of interaction on the evolution stage.

    Despite its utility, Method 2 has some limitations. The bi-mode fit's accuracy is questionable for peaks with few particles, and variations in fitting parameters can significantly alter the outcomes. Lastly, for interactions below $a_{dd}=1000a_0$, the fitting results are comparable with Method 1, except for the absence of error bars for individual data points. 

    For a more tangible perspective of interaction-induced damping, we plot the variation in the full width at half maximum (FWHM) of the $0\hbar k$ molecular clusters in Fig. \ref{fig:decay}(c). For demonstration, the timeline has been simplified to $N_{\rm evo} = 2\tau/T_{\rm actual}$, in which $T_{\rm actual}$ represents the actual Talbot time $T_{\rm T}$ under various interactions. An augmented FWHM suggests a wider momentum distribution and reduced coherence. The FWHM escalates with higher interaction strengths, notably surpassing $2\hbar k$ ($168\ \mathrm{\mu m}$) when $N_{evo} = 11$ at $a_{dd}=2004a_0$. Correspondingly, $\rm P_0$ values indicate an almost complete absence of condensed particles at zero momentum, signifying the interferometer's breakdown.

    \section{The interaction-induced Talbot period shift}\label{sec:shiftResult}
\begin{figure}
	\includegraphics[width=0.45\textwidth]{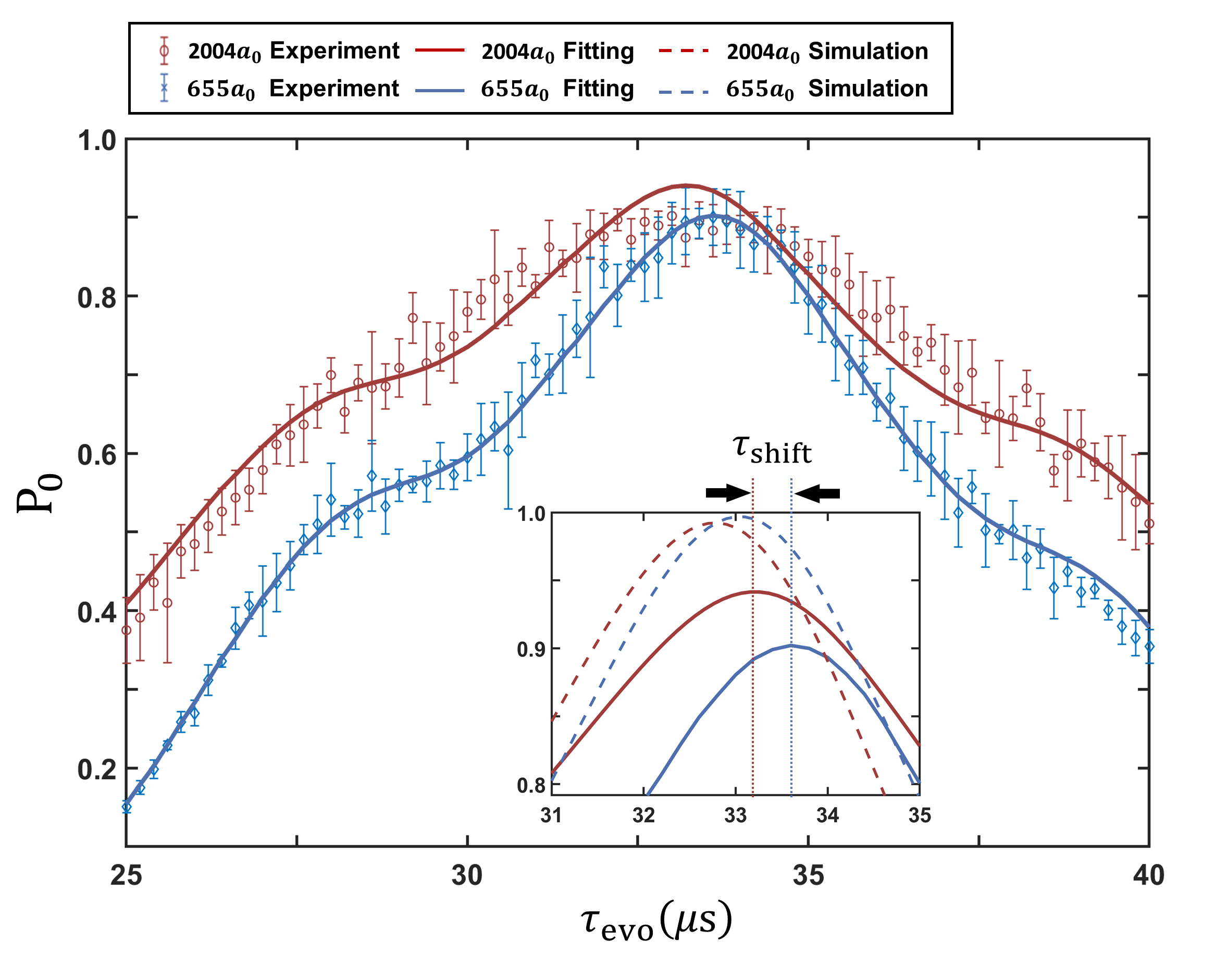}
	\caption{
		{\bf The shift of Talbot time caused by interactions}.
		The blue diamond dots and the red round dots show the experimental results of the oscillation of Talbot signals near $\tau = 3T_{\rm T}/2$ under $655 a_0$ and $2004a_0$, fitted with the blue solid line and red solid line, respectively. Details near the peak are shown in the inset, with the simulation results marked by the blue dashed line ($655 a_0$) and red dashed line ($2004 a_0$). }
	\label{fig:shift}
\end{figure}

    To investigate the shift of the Talbot period due to varying interaction strengths, we conduct experiments within a fixed $\tau$ region, specifically around $3T_{\rm T}/2$. This region is chosen because it accentuates the period shift while maintaining a relatively small decay of the signal. The selected time range is approximately one times $T_{\rm T}$. Then the experimental data are fitted with a no-interaction theoretical curve. The fitting process is guided by several parameters, which include the time scaling factor $r$, signal oscillation amplitude $A$, signal offset $d$, and decay constant $\tau_{\rm decay}$. The fitting expression is as follows:
    \begin{equation}\label{fit}
    \begin{aligned}
    &P_0= f( \tau_{\rm evo} ) \longrightarrow
    \\&{P}'_0= A f( (1+r)\tau_{\rm evo})\cdot e^{-\frac{\tau_{\rm evo}}{\tau_{\rm decay}} } 
    \\
    &+{A}' \left[1-f( (1+r)\tau_{\rm evo})\cdot e^{-\frac{\tau_{\rm evo}}{\tau_{\rm decay}} }\right]  + b,  
    \end{aligned}
    \end{equation}
    in which $0\le A+b\le1$, $0\le {A}'\le A$, $b\ge0$. The experimental results and corresponding fitting curves for interaction strengths of $a_{dd}=655a_0$ and $2004a_0$ are depicted in Fig. \ref{fig:shift}. We select the result with the smallest error after $100$ iterations of the fitting. Within the measured domain, the fitting curves are in good agreement with the experimental data points. A detailed view of the two curves around their respective maxima is presented in the inset, and the results of theoretical simulations are shown by the blue and red dashed lines. Table 1 presents the simulated and experimental values of $\tau_{\rm evo}$ aligned with the peaks under different interaction intensities.
\begin{center}
	TABLE I. The temporal shift of $\tau_{\rm evo} \sim 3T_{\rm T}/2$ under different interactions.
	\setlength\tabcolsep{3pt}
	\begin{tabular}{clll}
		\hline
		$a_{dd}$ ($a_0$)& $\mu/\hbar \bar{\omega }$ &Simulation ($\mathrm{\mu s}$) &Experiment ($\mathrm{\mu s}$)\\ 
		\hline
		655 &11.47&33.043 (0)&33.626 (0)\\
		865 &14.32&32.990 (-0.053)&33.527 (-0.099)\\
		1330 &19.45&32.904 (-0.139)&33.394 (-0.232)\\
		2004 &25.28&32.777 (-0.266)&33.082 (-0.544)\\
		\hline
	\end{tabular}
	\label{Table shift}
\end{center}
    
    The parentheses indicate the $\tau_{\rm shift}$ under various interactions compared to those obtained under $a_{dd}=655a_0$. The experimental and simulated results exhibit similar trends, with the values of $\tau_{\rm shift}$ also being relatively close. This change in $T_{\rm T}$, as described in Sec. \ref{sec:theoretic}, originates from the interaction's impact on $E_r$ or $U_0 \tau_0$ of optical lattices like an extra lattice. 
    
    Due to the limitations of the step size adjustment in the interference duration and the measurement errors near the maximum value, in experiments, it is only possible to qualitatively picture that increasing the interaction will shorten the Talbot time and cause a temporal peak shift. For quantitative measurements, as described in Sec. \ref{sec:B}, the accuracy can be enhanced by applying a pair of small-angle lattice beams with smaller $E_r$. 
    
    \section{Fractional Talbot effect}\label{sec:fractional}
\begin{figure}
	\includegraphics[width=0.45\textwidth]{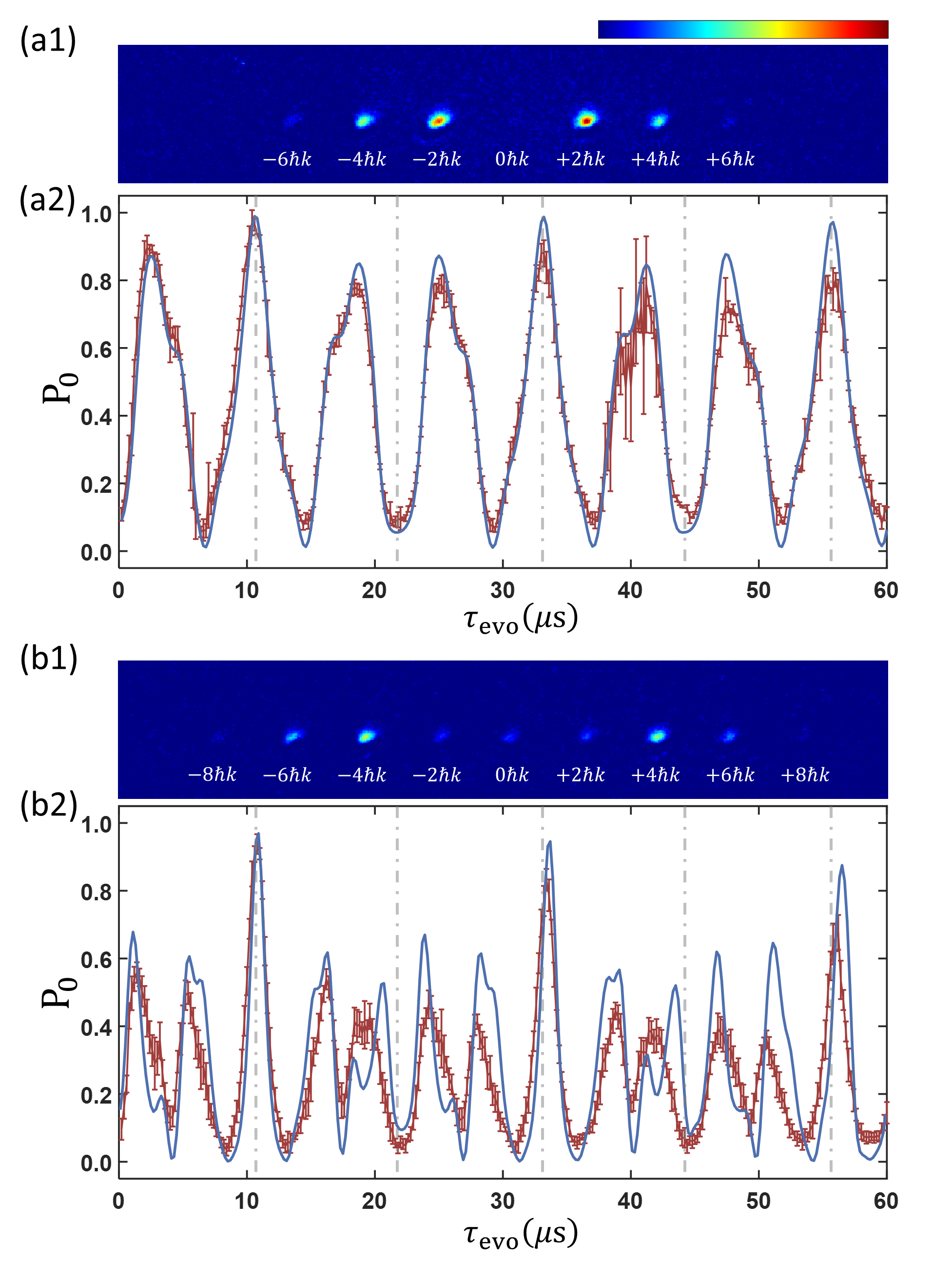}
	\caption{
		{\bf Fractional Talbot effect of strong interacting mBEC}.
		(a)$U_0=100E_r$. (b)$U_0=150E_r$. (a1)(b1) The initial state after the first lattice pulse ($\rm T_1$). (a2)(b2) Fractional Talbot interferometer with $a_{dd}=865 a_0$. The theoretical curves are marked by solid blue lines and the experimental data are marked by solid red dots with error bars. $\tau_{\rm evo}$ corresponding to integer multiples of $T_{\rm T}/2$ are depicted by gray dashed lines.}
	\label{fig:fractional}
\end{figure}
    Considering the fractional Talbot interferometer \cite{PhysRevLett.83.5407,Ullah2012} with higher-order momentum modes, collisions between particles with different momenta will become more intense during the interference process. To check the effectiveness under strong interactions, we modify the energy imparted to the particles by altering the initial and final pulses. We present the observation at different trap depths ($U_0$) and fixed pulse duration under the interaction strength of $a_{dd}=865 a_0$ ($691$ Gauss). The trap depth reaches values of $100 E_r$ and $150 E_r$, giving a initial states ($\rm T_1$) of Fig. \ref{fig:fractional}(a1) and (b1). For a trap depth of $U_0=50E_r$, the peak of $\rm P_0$ only appears when the evolution time equals to odd multiple of the half Talbot period. However, as illustrated in Fig. \ref{fig:fractional}(a2), for a trap depth of $U_0=100E_r$, in addition to the maximum at odd multiples of $T_{\rm T}/2$, there are two sub-maxima appearing between two adjacent peaks, symmetrically distributed on both sides of the maximum peak. When the optical lattice depth is increased to $U_0=150E_r$, even more sub-maxima emerge, as illustrated in Fig. \ref{fig:fractional}(b2). This observation is also confirmed by numerical simulations, showing the same results. The theoretical curves are marked by solid blue lines and the experimental data are marked by solid orange dots with error bars, in both Fig. \ref{fig:fractional}(a2)(b2).
    
    By comparing the initial states at different lattice depths, it can be observed that when the depths are $U_0=50E_r, 100E_r, 150E_r$ respectively, the dominant momentum in the initial state is $0\hbar k$, $\pm2\hbar k$, and $\pm4\hbar k$, while the number of maxima appearing in each Talbot period is 1, 3, and 5. Thus we can conclude that this fractional Talbot effect is caused by the interference among higher-order momentum modes. Furthermore, for initial states dominated by $\pm2n\hbar k$, there will be $2n+1$ maxima in each Talbot period. This conclusion is also supported by the results of GPE simulations.
    
    The influence of interactions on the fractional Talbot interferometer can be observed through a comparison of experimental and theoretical curves. In addition to the previously mentioned decay, there are also changes in the shape of certain peaks. As for the signal decay, the changes in peak values at odd multiples of $T_{\rm T}/2$ are not significantly different from those at low lattice depths. However, the decay of other sub-maxima is much faster, because the sub-maxima are mainly generated by higher-order momentum interference, and the dephasing caused by collisions between higher-order momentum modes is more pronounced as the interaction strength increases compared to lower-order momentum modes. The changes in peak shape arise from the slight modifications of the lattice depth due to interactions, which result in deviations between the experimental and theoretical initial states. Overall, the behavior of the fractional Talbot interferometer under strong interactions does not exhibit significant differences compared to the regular case.

    \section{Conclusion}\label{sec:Conclusion}

    In this study, we explore the effects of interaction on the Talbot signal's damping and temporal shift in a one-dimensional optical lattice. Our findings indicate that interactions minimally influence the Talbot signal during the pulse sequence. Conversely, intensified interactions in the evolution stage accelerate the damping of the signal markedly. This is also reflected by the widening response in the half-width at half-maximum of the $0 \hbar k$ momentum modes. Additionally, interactions induce a slight drift in the Talbot time $T_{\rm T}$, resulting in an earlier peak occurrence. However, within the experimentally acceptable interaction range, this shift in Talbot time is practically inconsequential. 
    
    Across a wide range of interactions, the Talbot interferometer remains highly effective over a certain evolutionary timescale, inclusive of fractional Talbot interference scenarios. Under the modification of theoretical quantification, it can be utilized for lattice parameter calibration, momentum filtering, and coherence measurement in strongly interacting systems.
    
    This work provides insight into the interplay between interaction and the coherence properties of a temporal Talbot interferometer in optical lattices, paving the way for research into quantum interference in strongly interacting systems.

	\section*{Acknowledgements}
	The authors thank Wenlan Chen and Xiaopeng Li for helpful suggestions in building this setup. This work is supported by the National Natural Science Foundation of China (Grants No. 92365208, No.11934002, and No. 11920101004), National Key Research and Development Program of China (Grants No. 2021YFA0718300 and No. 2021YFA1400900); C.L. is supported by the Austrian Science Fund (FWF) through the ESPRIT grant 10.55776/ESP310 (EAPQuP).

    \appendix
    \addcontentsline{toc}{section}{Appendices}\markboth{APPENDICES}{}
    \begin{subappendices}
		
    \section{Experimental Apparatus}\label{sec:A}

\begin{figure}
	\includegraphics[width=0.45\textwidth]{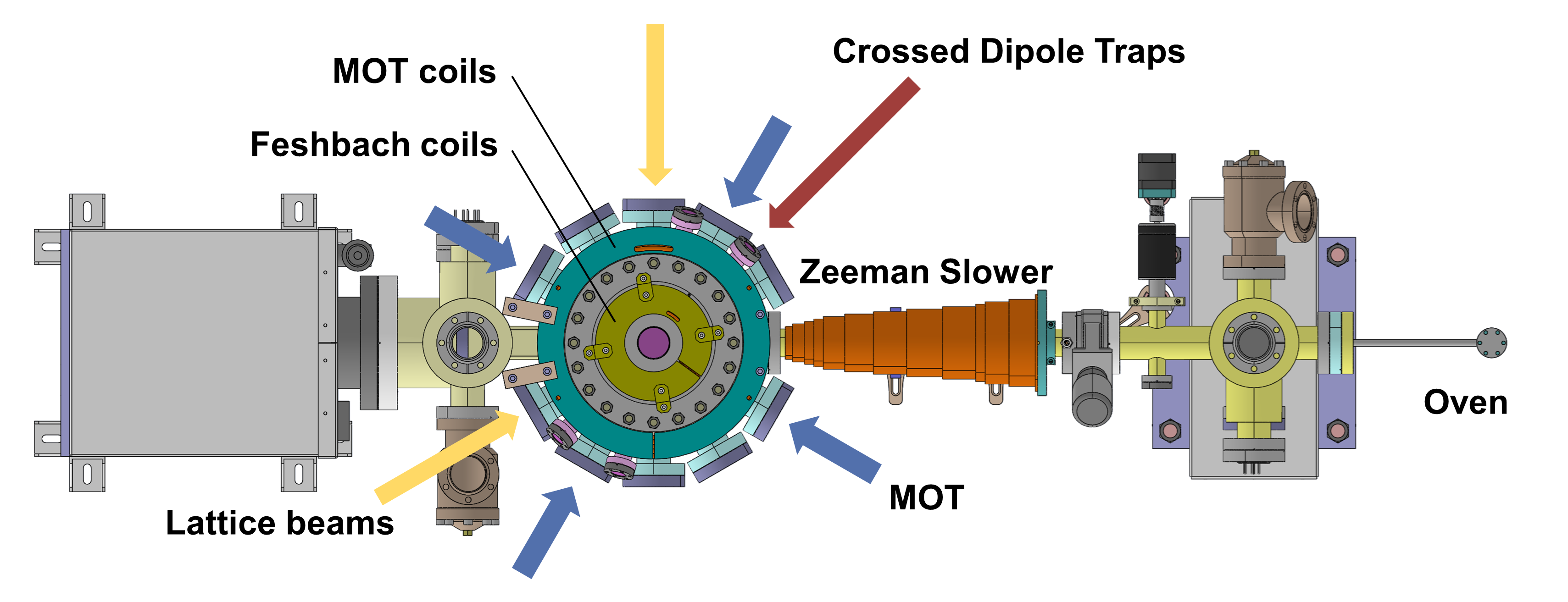}
	\caption{
		{\bf Experimental Apparatus}. Schematics of the experimental setup.}
	\label{fig:A}
\end{figure}

    The experimental setup, as illustrated in Fig. \ref{fig:A}, consists of an ultrahigh vacuum system comprising an atomic vapor oven, a Zeeman slower, and a science chamber. The vacuum pressure is maintained at an extremely low level of approximately $10^{-10} mbar$ in the oven section and a few $10^{-11} \rm mbar$ in the experimental chambers, thanks to the presence of two titanium sublimation pumps and two ion getter pumps ($40 \ \rm L$/$150\ \rm L$). Laser cooling is employed in three sequential steps: the magneto-optical trap (MOT), compressed magneto-optical trap (C-MOT), and gray molasses. The MOT collects lithium atoms, which have been laser cooled and collected from an oven through a Zeeman slower, at the experimental chamber using the $\rm D2$ line optical transition ($2^2s_{1/2} \to 2^2p_{3/2}$). The MOT comprises cooling and repump lights that excite atoms from the $F=3/2$ and $F=1/2$ states, respectively, and typically captures around $2\times10^9$ atoms at a temperature of approximately $1.5\ \mathrm{mK}$. The cooling and repumping beams are combined into the same optical fiber, which then generates three pairs of laser beams with a retro-reflection configuration. By ramping the laser frequency close to resonance and decreasing the optical intensity in the C-MOT, the temperature is further reduced to approximately $300\ \rm \mu K$, while maintaining around $8\times10^8$ atoms. The loading of atoms into the MOT takes around $8$ seconds, and the C-MOT lasts for about $35$ ms. In our experiment, the cooling and repumping light for the gray molasses has a blue detuning of approximately $6\Gamma$ from the $D1$ line transition ($2^2s_{1/2} \to 2^2p_{1/2}$), where the natural linewidth $\Gamma$ of the excited state is $5.87$ MHz. The laser beams for the gray molasses overlap with those of the MOT, facilitating optical alignment. The stage of gray molasses lasts for $2 \ \rm ms$ after switching off the magnetic field of the C-MOT, resulting in a reduction of the atomic temperature to $80\ \rm \mu K$, approximately one order of magnitude smaller.

    The creation of a Bose-Einstein condensate (BEC) necessitates the application of various experimental techniques, as well as the provision of an ultrahigh vacuum environment and stable laser light for atom trapping and imaging. For $\rm ^{6}Li$, the creation of homogeneous magnetic fields to form bosonic molecules that can be condensed is an additional requirement. After transferring the atoms to the optical dipole trap (ODT), the quadrupole magnetic field of the MOT is switched off, and a pair of Helmholtz coils provides homogeneous Feshbach magnetic fields.

    To achieve degenerate Fermi gases, we load cold atoms into an ODT for evaporative cooling. The ODT light is generated by a single-mode ytterbium-doped IPG fiber laser (YLR-$200$-$1064$-LP-WC). The ODT light is turned on $200$ ms before the end of the MOT. By extinguishing the repumping light of the gray molasses $100\ \rm \mu s$ earlier than the cooling light, the atoms are pumped to the $F=1/2$ states, which are the lowest two magnetic sublevels $|1\rangle $ and $|2\rangle $. The configuration of the ODT is shown in Fig. \ref {fig:experimentalSetup}(a), with two laser beams focused and intersecting in the science chamber at an angle of $30^{\circ}$. The waist radius of the laser beam is $34\ \rm\mu m$. To avoid optical interference, two acousto-optical modulators (AOMs) are used to control the laser beams at frequencies of $+110$ MHz and $-110$ MHz, respectively. Evaporation cooling is then performed by decreasing the optical power of the dipole trap. The evaporation process is carried out under a magnetic field offset of $832$ Gauss, where the s-wave scattering length is infinitely large, resulting in strong interaction between the spin states and rapid thermalization. The ODT light is initially turned on with a power of $200$ W for each beam and kept on for approximately $200\ \rm ms$ to reach equilibrium. The laser power is then ramped down through a two-stage exponential attenuation scanning process with a period of approximately $1.8\ \mathrm{s}$. The percentage of laser power is monitored by a photodetector. In the first stages, the laser power is controlled by an external voltage, while in the final stages, a PI locking circuit is introduced to stabilize the optical intensity. As the laser power decreases, the atomic temperature also decreases. When the laser power is further ramped down to approximately $10\ \mathrm{mW}$, the Fermi gas becomes degenerate, with $T/T_{\rm F} \approx 0.1$, where $T$ is the atomic temperature and $T_{\rm F}$ is the Fermi temperature of the non-interacting Fermi gas.

    \section{Calculation of Interacting Talbot Interferometers}\label{sec:B}
\begin{figure}
	\includegraphics[width=0.45\textwidth]{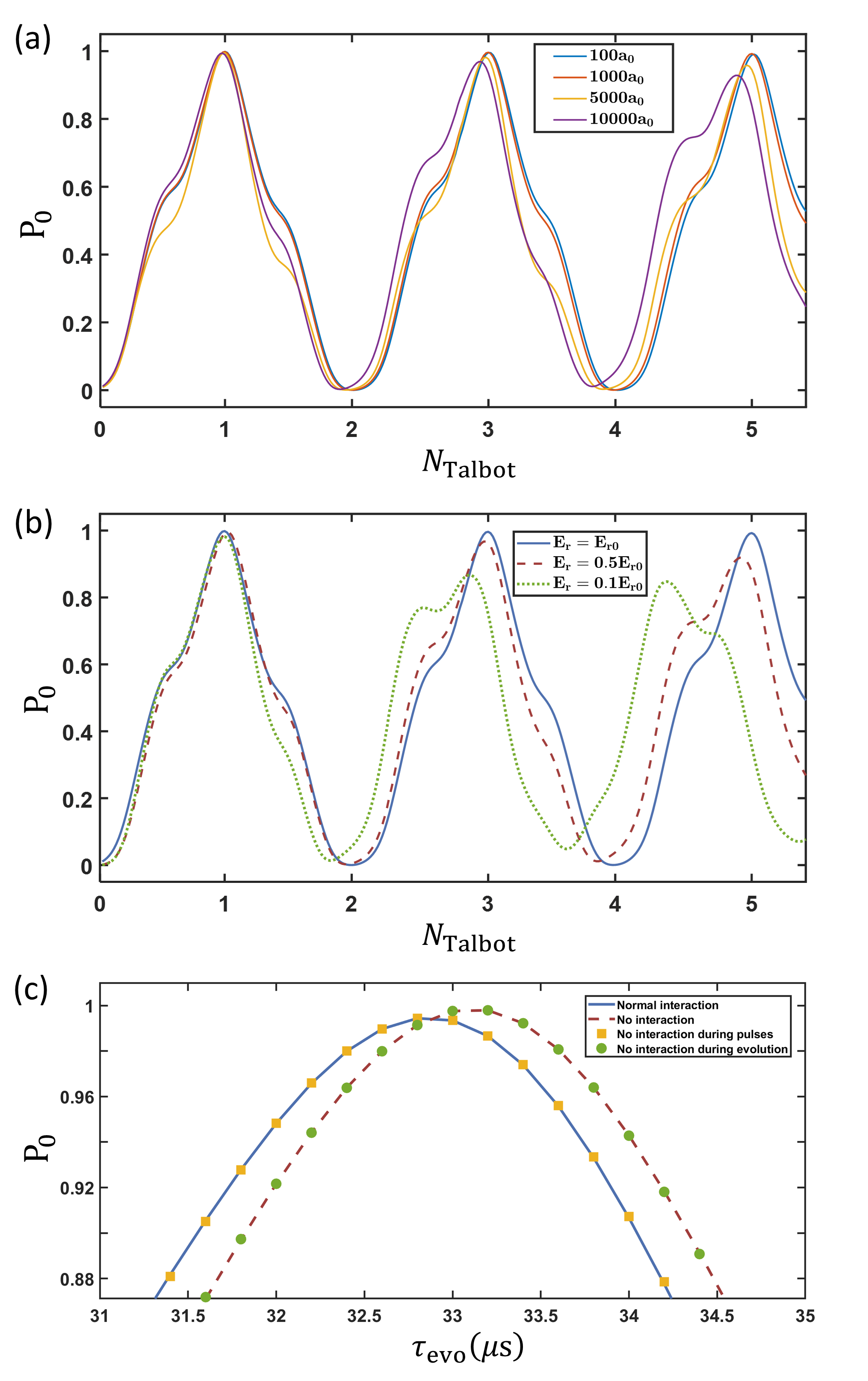}
	\caption{
		{\bf Theoretical calculation of strongly interacting Talbot interferometer}. The optical lattice pulse pattern is the same with lattice depth $U_0 = 50E_{r0}$ and pulse duration $\tau_0 = 0.7\ \rm{\mu} s$. The particle number is set as 30000. (a) Talbot interferometer at different scattering lengths. The blue line, red line, yellow line, and purple line represent the simulation results with $a_{dd}= 100a_0$, $1000a_0$, $5000a_0$ and $10000a_0$ respectively. $N_{\rm Talbot} = 2\tau/T_{\rm theoretic}$. (b) Talbot interferometer at $a_{dd} = 1000 a_{0}$ with different characteristic lattice energy $E_r$. The blue line, red dotted line, and green dotted line represent the simulation results with $E_r=E_{r0}$, $E_r=0.5E_{r0}$ and $E_r=0.1E_{r0}$ respectively. (c) The results of simulations with different interactions during the pulse stage and evolution stage. The blue solid line represents the results with constant interaction $a_{dd}= 2004a_0$, while the red dashed line indicates scenarios where no interaction occurs throughout. The yellow squares and green circles denote the results obtained when interaction is absent during the pulse phase and evolution phase, respectively.}
	\label{fig:simulation}
\end{figure}    
\begin{figure}
	\includegraphics[width=0.45\textwidth]{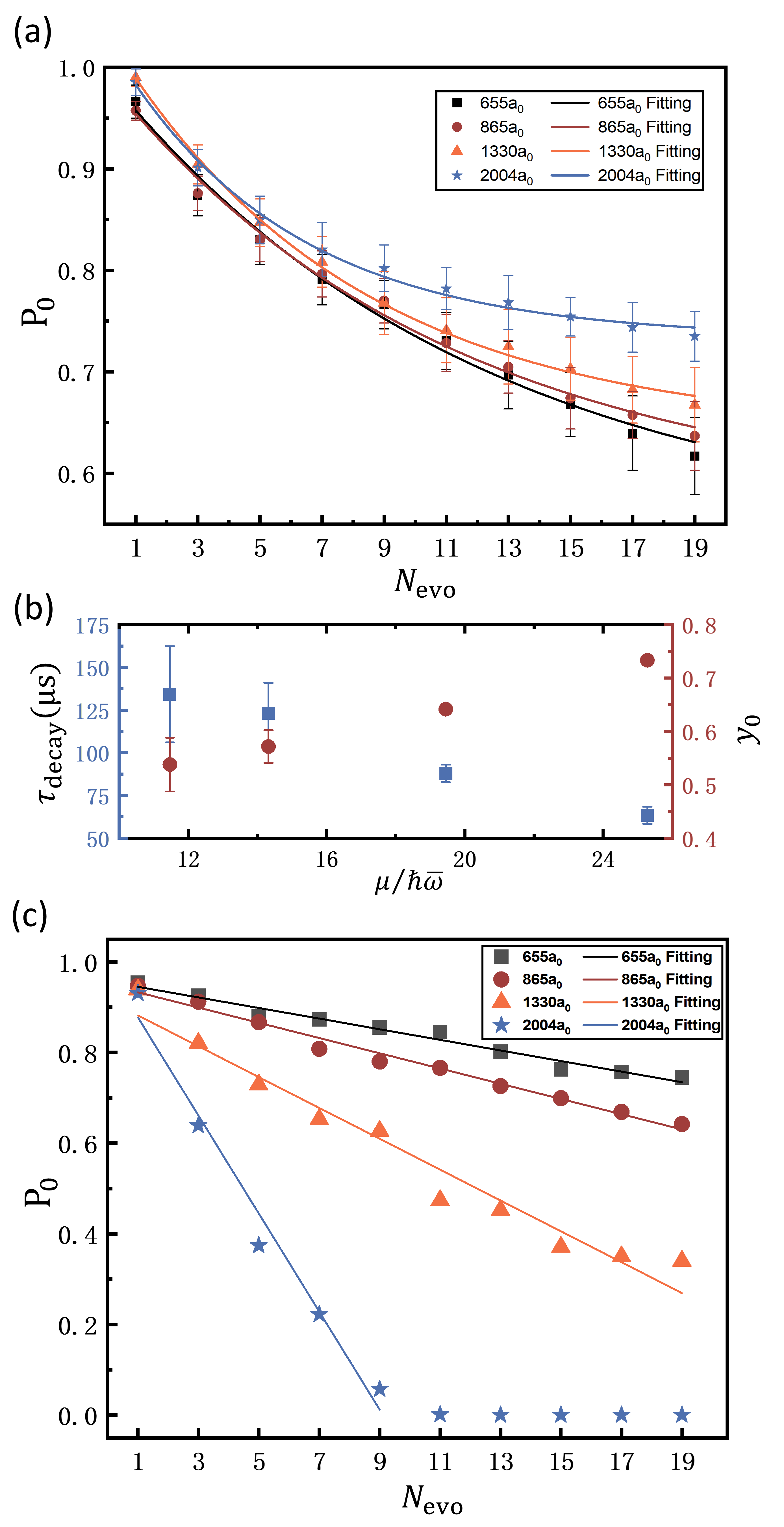}
	\caption{
		{\bf Talbot signal decay in Method 1 and Method 2.} 
        (a) The decay curve obtained with Method 1 under different interactions. The black squares, red circles, orange triangles, and blue pentagons represent the experimental results of $a_{dd}=655 a_0$, $865 a_0$, $1330 a_0$ and $2004 a_0$, respectively. The fitting results under each interaction strength are indicated by lines of the corresponding colors. (b) The exponential fitting parameters for the decay curves under different interactions. $\tau_{\rm decay}$ is marked by the blue square dots, and $y_0$ the red round dots. (c) The decay curve is obtained with Method 2 under different interactions. The black squares, red circles, orange triangles, and blue pentagons represent the experimental results of $a_{dd}=655 a_0$, $865 a_0$, $1330 a_0$ and $2004 a_0$, respectively. The linear fitting results under each interaction strength are indicated by lines of the corresponding colors.}
	\label{fig:C}
\end{figure}
    To understand the effect of interaction in Talbot interference, we implemented (i) simulation at different scattering length with same optical lattice pulses ($U_0 = 50 E_r$, $\tau_0 = 0.7 $$\rm{\mu s}$) and (ii) simulation at scattering length $a_{dd} = 1000 a_{0}$ with same optical lattice pulses ($U_0 = 50 E_r$, $\tau_0 = 0.7 $$\rm{\mu s}$) but different characteristic lattice energies.
    
   Fig. \ref{fig:simulation} (a) shows the results of (i), the Talbot effect with different interaction strengths. We observe increasing negative shifts of the peak in height and position with increasing scattering length. With a larger scattering length, the effect of interaction becomes significant so that the profile of the curve has slightly changed. Fig. \ref{fig:simulation} (b) shows the results of (ii), Talbot effect with the same interaction strength and optical lattice pulses for small characteristic lattice energy $E_r$ (larger lattice spatial depth $D$). With smaller values of $E_r$, the Talbot effect shows similar changes in interaction as illustrated in Fig. \ref{fig:simulation}(a), while the Talbot effect with no interaction remains unaffected. It is easy to understand because the decrease of $E_r$ equals the increase in interaction.
    
    To distinguish the effect of interaction during the pulse stage and free evolution stage, we perform simulations by tuning interactions during these two stages separately. As demonstrated in Fig. \ref{fig:simulation}(c), when the pulse stage is devoid of interaction, the results are indistinguishable from those where interactions are present throughout the entire interference process. Conversely, when the evolution stage lacks interactions, the results align with those obtained in the absence of any interactions during the entire sequence. It indicates that the influence of interaction focuses more on the free evolution stage. It is predictable because the lattice trap depth is too large for the interaction and the duration is too short for the free evolution time.

\section{Decay Sequences}\label{sec:C}
    In Sec. \ref{sec:decayResult}, due to the disturbance caused by thermal particles near $0\hbar k$ region in the Talbot signals obtained by the traditional method (Method 1) under strong interactions, we introduce Method 2 as the basis for judging signal damping. As shown in Fig. \ref{fig:C}(a), the Talbot signal damping obtained through Method 1 under different interactions shows no significant relevance with interaction strengths. Even in the case of the highest interaction strength ($a_{dd}=2004 a_0$), the decay curve appears the smoothest.
    
    However, although the changes in signal damping observed are not intuitive, we can still obtain the effects of interactions from their fitting parameters. Considering the most common exponential fitting: $f( \tau_{\rm evo} ) = A\cdot e^{-\frac{\tau_{\rm evo}}{\tau_{\rm decay}} }+ y_0$, we extracted $\tau_{\rm evo}$ and $y_0$ under different interactions and presented them in Fig. \ref{fig:C}(b). $y_0$ increases with higher interaction strengths, indicating that the interaction promotes the signal damping to approach a higher $P_0$, which is in good agreement with our judgment of the influence of thermal particles generated during the interference period. Meanwhile, higher interaction strengths bring smaller $\tau_{\rm evo}$, showing a shorter decay characteristic time scale, which agrees well with the decay constants obtained through Method 2. In summary, the experimental results obtained through Method 1 and Method 2 can both confirm the theoretical prediction that interaction leads to faster Talbot signal damping.
    
    The decay curves obtained with Method 2 under different interactions are shown in Fig. \ref{fig:C}(c). When we only include condensate parts in the calculation of the Talbot signal, the contrast of signal damping under different interactions is much better than the signal obtained from Method 1. When the interaction strength is below $a_{dd}=865 a_0$, the signal damping is very slow, even slower than the results obtained from Method 1. When the interaction strength is above $a_{dd}=1330 a_0$, the decay significantly accelerates, and at $a_{dd}=2004 a_0$, the proportion of condensate in the $0\hbar k$ mode is almost zero. This is because the width of the thermal particle component remains almost constant, contributing to both the $0\hbar k$ mode and the other higher modes ($\pm 2\hbar k$, $\pm 4\hbar k$) regardless of the interaction. When the interaction strength is relatively weak, the optical density of the thermal particles is relatively small, and the contribution to the $0\hbar k$ mode is smaller (in proportion) compared to the other higher modes, thus reducing the signal statistically. On the contrary, when the interaction strength is strong. Method 2 eliminates this disturbance and can therefore obtain a clearer Talbot signal variation curve. As for why the signal attenuation shows good linearity, further research is needed.

    \end{subappendices}


\bibliographystyle{apsrev}
\bibliography{ref}

 
\end{document}